\documentclass[pre,twocolumn,showpacs,superscriptaddress,floatfix]{revtex4}
\newcount\tipo\tipo=1 
\usepackage{amsmath}
\usepackage{graphicx}

\let\a=\alpha\let\b=\beta  \let\d=\delta
 \let\z=\zeta 
\let\k=\kappa \let\l=\lambda \let\m=\mu 
  \let\r=\rho \let\s=\sigma \let\t=\tau
 \let\f=\varphi  
  \let\D=\Delta 
 \let\P=\Pi  \let\F=\Phi
 \let\O=\Omega 

\let\dpr=\partial\let\0=\noindent\let\fra=\frac
\def\media#1{\langle{#1}\rangle}\let\==\equiv
\def\*{\vskip3mm}\def\eg{{\it e.g. }}\def\ie{{\it i.e.\ }}
\def\V#1{{\bf#1}}\let\io=\infty\let\ig=\int
\def\otto{\,{\kern-1.truept\leftarrow\kern-5.truept\to\kern-1.truept}\,}
\let\wt=\widetilde\def\Defi{\,{\buildrel def\over=}\,}

\def\Onlinecite#1{[\onlinecite{#1}]}
\def\tende#1{\ \vtop{\ialign{##\crcr\rightarrowfill\crcr
\noalign{\kern-1pt\nointerlineskip} \hglue3.pt${\scriptstyle%
#1}$\hglue3.pt\crcr}}\,}
\newcommand\defi{\,{\buildrel def \over =}\,}
\newcommand\revtex{{R\kern-1mm\lower0.5mm\hbox{E}\kern-0.6mm V\kern-0.5mm%
\lower0.5mm\hbox{T}\kern-0.5mm E\kern-.5mm \lower0.5mm\hbox{X}}}
\begin{document}
\preprint{FM 04-04} 

\title{Chaotic Hypothesis, Fluctuation Theorem and Singularities}
\author{F.~Bonetto}
\affiliation{School of Mathematics, 
Georgia Institute of Technology, Atlanta Georgia 30332}
\affiliation{Dipartimento di Matematica, Universit\`a di Roma 
{\em Tor Vergata}, 
V.le della Ricerca Scientifica, 00133, Roma, Italy}
\author{G.~Gallavotti}\affiliation{INFN, Universit\`a di Roma 
{\em La Sapienza},
P.~A.~Moro 2, 00185, Roma, Italy}
\affiliation{Dipartimento di Fisica, Universit\`a di Roma 
{\em La Sapienza},
P.~A.~Moro 2, 00185, Roma, Italy}
\author{A.~Giuliani}
\affiliation{Dipartimento di Matematica, Universit\`a di Roma 
{\em Tor Vergata}, 
V.le della Ricerca Scientifica, 00133, Roma, Italy}
\affiliation{INFN, Universit\`a di Roma {\em La Sapienza},
P.~A.~Moro 2, 00185, Roma, Italy}
\author{F.~Zamponi}
\affiliation{Dipartimento di Fisica, Universit\`a di Roma {\em La Sapienza},
P.~A.~Moro 2, 00185, Roma, Italy}
\affiliation{SOFT-INFM-CNR, Universit\`a di 
Roma {\em La Sapienza},
P.~A.~Moro 2, 00185, Roma, Italy}
\relax 
\begin{abstract}
The chaotic hypothesis has several implications which have generated
interest in the literature because of their generality and because a few 
exact predictions are among them. 
However its application to Physics problems requires
attention and can lead to apparent
inconsistencies.  In particular there are several cases that have
been considered in the literature in which singularities are built in
the models: for instance when among the forces there are Lennard-Jones
potentials (which are infinite in the origin)
and the constraints imposed on the system do not forbid
arbitrarily close approach to the singularity even though the average
kinetic energy is bounded. 
The situation is well understood in certain
special cases in which the system is subject to Gaussian noise; here
the treatment of rather general singular systems is considered and
the predictions of the chaotic hypothesis for such situations are
derived. The main conclusion is that the chaotic hypothesis is perfectly
adequate to describe the singular physical systems we consider,
\ie deterministic systems with thermostat forces acting according to
Gauss' principle for the constraint of constant total 
kinetic energy (``isokinetic Gaussian thermostats''), close and 
far from equilibrium.
Near equilibrium it even predicts a fluctuation
relation which, in deterministic cases with more general  
thermostat forces (\ie
not necessarily of Gaussian isokinetic nature), extends recent
relations obtained in situations in which the thermostatting forces
satisfy 
Gauss' principle. This relation agrees, where expected, with the fluctuation
theorem for perfectly chaotic systems. The results are compared with
some recent works in the literature.
\end{abstract}
\maketitle

\section*{Introduction}

There is a quite strong interest in stationary states of systems
subject to the action of non conservative forces. These forces perform work
on the system while by suitable mechanisms heat is extracted,
so that the system can stay in a statistically stationary state.
Theoretical and experimental works are steadily becoming avalaible on
the matter. Theoretical work implement the heat extraction in several
ways introducing ``thermostat models'', which can be stochastic or
deterministic forces.

A strong idealization of a system in a nonequilibrium steady state
subject to deterministic forces is provided by the Anosov systems:
their motion can be considered to be paradigm of chaotic behavior,
playing in chaotic dynamics the role that harmonic motions play in
regular dynamics. The chaoticity of the motions is immediately
apparent from the definition of Anosov systems: locally around each
point it has to be possible to draw three coordinate surfaces
$W_s,W_u,W$ such that segments of curves on $W_s,W_u$ contract
exponentially as time grows to $+\infty$ or, respectively, recedes to
$-\infty$ while segments on $W$, the one dimensional ``neutral'' flow
direction, neither expand nor contract. They change their length but keep
it of the same order of the initial one. If the dynamics is
described by a map the neutral direction is omitted in the
definition.

The chaotic hypothesis, \cite{GC95,Ru99}, see below, proposes 
that chaotic systems
should be considered as Anosov systems ``for practical
purposes''. This has several consequences: in particular about
fluctuations in time reversible models, where the hypothesis
leads to severe constraints through the {\it Fluctuation Theorem}.
This is a mathematical property of the large deviations function of the
phase space contraction of a {\it time reversible} Anosov map $S$.
However, some obvious restrictions, analogous to the ones that are 
(often tacitly) assumed
when one says that the ``pendulum is isochronous'' or that phonons
in a crystal correspond to ``harmonic excitations'', have to be taken into
account when applying the hypothesis to realistic systems, which are not
strictly Anosov systems.

The prediction has been tested in several simulations and we summarize
the precise statement of it below: usually the results have been
positive. 
However, there have been, in the
literature, a few claims of failure of the chaotic hypothesis 
based on the apparent failure of
the predictions of the fluctuation theorem.
Here we concentrate on one such attempt, which studies
systems violating the Anosov property because singularities of the
interparticle potentials play an important role in the dynamics,
\cite{ESR03}: a situation considered, correctly, in the literature as
not important for most physical properties but which requires care if
the fluctuation relation is specifically tested on such systems (in
the same way care has to be used if isochrony is tested on a pendulum
or harmonicity is tested in a crystal model).

Here we show that even in singular systems the chaotic hypothesis and
the fluctuation theorem are not in contradiction: we develop a theory
that extends the fluctuation theorem to singular systems continuing
ideas that were introduced to study a special Gaussian noise
thermostat,~\cite{CV03a}, (this is a ``random thermostat'' not to be
confused with the thermostats satisfying Gauss' principle for some non
holonomic constraint, like the isokinetic constraint).

The structure of the paper is the following: in
section~\ref{sec:I} we recall the basic notations and statements, and
some alternative formulations of the fluctuation relation.
We discuss its (trivial) form in equilibrium and how one can take a
meaningful and non-trivial equilibrium limit.
In section~\ref{sec:II} we discuss the application of the chaotic
hypothesis to singular systems. First we present a very simple example
which shows that the effect of singularities is very important.
Then we discuss how one can obtain quantitative predictions on
the modification of the fluctuation relation due to the presence of
singularities. Finally we discuss a prescription to remove singularities
that follows from a careful examination of the proof of the fluctuation
theorem for Anosov flows. The results are compared with recent numerical
simulations. In section~\ref{sec:III} we draw the conclusions and compare
our interpretation with the one of \cite{ESR03}.

\section{The fluctuation relation}
\label{sec:I}

We shall denote by $\O$ the {\it phase space} (a smooth compact boundaryless
Riemannian manifold), by $S:\O\to\O$ an invertible map on $\O$
and by $\s(x)$ the volume {\it contraction}
\begin{equation}\s(x)=-\log |\det \dpr_xS(x)|\label{1}\end{equation}
{\it Time reversal} is defined as an isometry $I: \O\otto\O$ with
\begin{equation}I S = S^{-1}I, \qquad \s(Ix)=-\s(x)\label{2}\end{equation}
If $S$ is an Anosov maps, 
existence of a unique invariant probability distribution
$\m$, called the {\it SRB distribution} and 
describing the long--time statistics of the motions whose initial data
are chosen randomly with respect to the volume measure, 
is established, \cite{Ru95,GBG04}. It has the property that, with the
exception of points $x\in\O$ in a set of $0$--volume, we have
\begin{equation}\lim_{\t\to\infty} \fra1\t \sum_{t=0}^{\t-1} F(S^tx)\defi
\media{F}=\int_\O F(y)\m(dy)\label{3}\end{equation}
for all smooth observables $F$ defined on phase space.

It is intuitive that ``phase space cannot expand''; this is expressed
by the following result of Ruelle \Onlinecite{Ru96}:

\*
{\it If $\s_+\defi\media{\s}$ it is $\s_+\ge0$} 
\*

Clearly if $S$ is volume preserving $\s_+=0$. If $\s_+>0$ the
system does not admit any stationary distribution of the form
$\m(dx)=\r(x)dx$, with density with respect to the volume measure $dx$
(often called {\it absolutely continuous} with respect to the volume).

This motivates calling systems for which $\media{\s}>0$ {\it dissipative}
and {\it conservative} the others.

For Anosov systems which are {\it transitive} ({\it i.e.} with a dense orbit),
reversible and dissipative one can define the dimensionless
phase space contraction, a quantity often related to entropy creation rate
(see \Onlinecite{Ga04b}), averaged over a time interval of size $\t$.
This is
\begin{equation}p(x)= \fra1{\s_+ \t }\sum_{k=-\t/2}^{\t/2-1} 
\s(S^k x)\label{4}\end{equation}
provided {\it of course} $\s_+>0$. 

Then for such systems the probability with respect to the stationary
state, {\it i.e.} to the SRB distribution $\m$, that the variable $p(x)$ takes
values in $\D=[p,p+\d p]$ can be written as $\P_\t(\D)=e^{\t
\max_{p\in \D}\z(p) + O(1)}$, where $\z(p)$ is a suitable function and,
for any fixed choice of $\D$ 
contained in an open interval $(-p^*,p^*)$, $p^*\ge 1$, the correction 
term at the exponent is $O(1)$ with respect to $\t^{-1}$, 
as $\t\to\io$ (this is often informally expressed as 
$\lim_{\t\to\infty}\fra1\t 
\log \P_\t(p)=\z(p)$ for $\ -p^*<p<p^*$). 
The function $\z(p)$ is called in probability
theory the {\it rate function} for the large deviations of $p$.

The function $\z(p)$ is analytic in $p$ and {\it convex} in the interval of
definition $(-p^*,p^*)$. Analyticity and convexity
of large deviation rates are general properties, established by Sinai and
valid for the SRB-averages of smooth observables (in Anosov systems),
\cite{Si72,Si77,GBG04}. In fact more can be said for the specific case
of the large deviation rate
of the observable $p$, and one can prove the following 
{\it fluctuation theorem}:

\*{\it In transitive time reversible dissipative
Anosov systems the
rate function $\z(p)$ for the dimensionless phase space contraction $p(x)$ 
defined in (\ref{4}) is
analytic and strictly convex in an interval $(-p^*,p^*)$ with
$+\infty>p^*\ge1$ and $\z(p)=-\infty$ for $|p|> p^*$.  Furthermore
\begin{equation}\z(-p)=\z(p)-p\s_+, \qquad {\rm for} \qquad
  |p|<p^*\label{5}
\end{equation}
which is called the ``fluctuation relation'' (FR).
}
\*

Strict convexity follows from a theorem of Griffiths and Ruelle which
shows that the only way strict convexity could fail is if
$\s(x)=\f(Sx)-\f(x)+ c$ where $\f(x)$ is a smooth function (typically
a Lipschitz continuous function) and $c$ is a constant, see
propositions (6.4.2) and (6.4.3) in \Onlinecite{GBG04}. The constant $c$
vanishes if time reversal holds and $\s(x)=\f(Sx)-\f(x)$ contradicts
the assumption that $\s_+>0$, because $\t^{-1}\sum_{-\t/2}^{\t/2-1}
\s(S^k x)=\t^{-1}\big[\f(S^{\t/2-1}x)-\f(S^{-\t/2}x)\big]\to 0$ as
$\t\to\infty$.  

The value of $p^*$ must be $p^*\ge1$ otherwise the
average of $p$ could not be $1$ (as it is by its very definition):
it is defined, adopting the natural convention that $\z(p)=-\io$ 
for the values of $p$ whose
probability goes to $0$ with $\tau$ faster than exponentially,
as the infimum of the
$p>0$ for which $\z(p)=-\io$.
Alternatively $\pm p^*$ are the asymptotic slopes as
$\l\to\pm\io$ of the Laplace transform $\log \media{e^{\l p}}_{SRB}$ 
\cite{Ga95b}. 

The fluctuation relation was discovered in a numerical
experiment, \Onlinecite{ECM93}, dealing with a non smooth system
(hence not Anosov). The formulation and proof of the above proposition is in
\Onlinecite{GC95} and in the context of Anosov systems the relation (\ref{5})
is properly called the {\it fluctuation theorem}.
The difference between this theorem and other fluctuation relations
proposed in the literature has been clarified in \cite{CG99}.
The theorem can be extended to Anosov flows ({\it i.e.} to systems
evolving in continuous time), \cite{Ge98}.  \*

\subsection*{Alternative formulations}

Sometimes, {\it e.g.} in \cite{SE00,ESR03}, rather than the above $p$ the
quantity $a=\t^{-1}\sum_{j=-\t/2}^{\t/2-1} \s(S^j x)$ is considered
and eq.(\ref{5}) becomes
\begin{equation}\widetilde\z(-a)=\widetilde\z(a)-a,\qquad {\rm for}\ |a|< a^*\=
p^*\s_+\label{6}\end{equation}
where $\widetilde\z(a)$ is trivially related to $\z(p)$.  This form
dangerously suggests that in systems with $\s_+=0$ the
distribution of $a$ is asymmetric (because the extra
condition $|a|<p^*\s_+$ might be forgotten, see \Onlinecite{Ga04}).

Note that $p^*$ is certainly $<+\infty$ because the variable $\s(x)$
is bounded (being continuous on the bounded manifold on which the
Anosov map is defined).

However no confusion should be made between $p^*\s_+$ and
$\sigma_{max}\defi\max |\s(x)|$: unlike $\sigma_{max}$ the quantity
$p^*$ is a {\it non trivial} dynamical quantity, independent on the
metric used on phase space to measure distances, hence volume. 
This point has not been always understood and confusion has appeared 
in the published literature with unexpected consequences. 
In fact it is very easy to build examples of Anosov systems in which 
$p^*\s_+< \sigma_{max}$:
still, this does not mean that fluctuation relation is violated for
such systems. Some explicit examples are discussed in next Section.

\subsection*{Conservative systems and the equilibrium limit}

Considering more closely the cases $\s_+=0$ it follows that
$\s(x)=\f(Sx)-\f(x)$ (again by the above mentioned result of 
Griffiths and Ruelle), with $\f$ a smooth function of phase space. 
Hence the variable
\begin{equation}a=\fra1\t\sum_{j=-\t/2}^{\t/2-1}
\s(S^jx)\=\frac{\f(S^{-\frac{\t}2}x)-
\f(S^{\frac{\t}2}x)}\t
\label{7}\end{equation}
is {\it bounded} and tends to $0$ {\it uniformly}. One could repeat
the theory developed for $p$ when $\s_+>0$ but one would reach the
conclusion that $\widetilde \z(a)=-\infty$ for $|a|>0$ and we see that
the result is trivial. In fact in this case it follows that the
system admits an absolutely continuous SRB distribution. The
distribution of $a$ is symmetric (trivially by time reversal
symmetry) and becomes a delta function around $0$ as $\t\to\infty$.

Nevertheless the fluctuation relation is {\it non trivial} in cases in
which the map $S$ depends on parameters $\underline
E=(E_1,\ldots,E_n)$ and becomes volume preserving (``conservative'')
as $\underline E\to \underline0$: in this case $\s_+\to0$ as
$\underline E\to\underline 0$ and one has to rewrite the fluctuation
relation in an appropriate way to take a meaningful limit.

The result is that the limit as $\underline E\to\underline 0$ 
of the fluctuation relation
in which both sides are divided by $\underline E^2$ makes sense and yields (in
the case considered here of transitive Anosov dynamical systems)
relations which are non trivial and that can be interpreted as giving
Green--Kubo formulae and Onsager reciprocity for transport
coefficients, \Onlinecite{Ga96a,GR97}.

In fact the very definition of the duality between currents and fluxes
so familiar in nonequilibrium thermodynamics since Onsager can be set
up in such systems using as generating function the $\s_+$ regarded as
a function of $\underline E$. Note that the fluxes are usually ``currents''
divided by the temperature: therefore via the above interpretation one
can try to define the temperature even in nonequilibrium situations,
\Onlinecite{GC04,Ga04b,ZCK05}.

\section{Singular systems}
\label{sec:II}

The fluctuation relation has been proved only for Anosov 
systems.
However, a {\it Chaotic Hypothesis} has been proposed, which
states that, {\it for the purpose of studying the physically 
interesting observables}, 
a chaotic dynamical system can be considered
as an Anosov system, \cite{GC95,GC95b,Ga00}.

In applying the chaotic hypothesis to singular 
systems, \eg a system of particles interacting via a
Lennard-Jones potential (which is infinite in the origin),
one might encounter apparent difficulties. We will discuss them in
the following.

\subsection*{The effect of a change of metric for Anosov flows}

The simplest example (out of many) is provided by the simplest
conservative system which is strictly an Anosov transitive system and
which has therefore an SRB distribution: this is the geodesic flow $S_t$
on a surface of constant negative curvature, \Onlinecite{BGM98}.  We
discuss here an evolution in continuous time because the matter is
considered in the literature for such systems, \Onlinecite{Ga04} (even
simpler examples are possible for time evolution maps).

The phase space $M$ is compact, time reversal is just momentum
reversal and the natural metric, induced by the Lobatchevsky metric
$g_{ij}(q)$ on the surface, is time reversal invariant: the SRB
distribution is the Liouville distribution and $\s(x)\equiv0$. However
one can introduce a function $\F(x)$ on $M$ which is very large in a
small vicinity of a point $x_0$, arbitrarily selected, constant
outside a slightly larger vicinity of $x_0$ and positive everywhere. A
new metric could be defined as $g_{new}(x)=(\F(x)+\F(Ix)) g(x)\defi
e^{-F(x)} g(x)$: it is still time reversal invariant but its volume
elements {\it will no longer be invariant} under the time evolution
$S_t$ associated with the geodesic flow with respect to the
Lobatchevsky metric. The
rate of change of phase space volume in the new metric will be
$\sigma_{new}(x)=\frac{d}{dt}F(x)$. Then the
phase space contraction $\s_{new}(x)$ takes values that not only are
not identically $0$ but which can in general be arbitrarily large, 
depending on the specific choice of $\Phi(x)$. 
The distribution of $a=\fra1\t \int_0^\t
\s_{new}(S_t x) dt= \t^{-1} \big[F(S_\t x)-F(x)\big]$,
at any finite time, will violate (\ref{6}), simply because it is symmetric 
around $0$, by time reversal. 

In the limit $\t\to\io$, as long as $F(x)$ is bounded, 
$a= \t^{-1} \big[F(S_\t
x)-F(x)\big]\tende{\t\to\infty}0$ uniformly in $x$, 
as in the corresponding map case, and the SRB
distribution of $a$ will tend to a delta function centered in $0$ (hence
$\widetilde\z(a)=-\infty$ for $a\ne0$). However, if $F(x)$ is not bounded
(\eg if it is allowed to become infinite in $x_0$) this is not the case 
in general, as we shall discuss in detail in next section. 

\subsection*{The effect of singular boundary terms}

One can realize that terms of the form $\t^{-1}\big[F(S_\t x)-F(x)\big]$ with
$F(x)$ not bounded can affect the large fluctuations of $\s(x)$, at least if 
the probability of an arbitrarily large value of $F$ is not too small, \ie if
asymptotically for big values of $F$ it is exponentially small in $F$
(or larger), \eg it is
of the form $\sim e^{-\k F}$, for some constant $\k>0$. This
is a valuable and interesting remark brought up for the first time,
and correctly interpreted, already in \cite{CV03a} and in the following papers
\cite{CV03,VCC04}. The analysis of \cite{CV03a,CV03,VCC04} applies
to cases where the unbounded fluctuations are driven by an external 
white noise. In the following we extend
the theoretical analysis in \cite{CV03,VCC04} to cases in which the
unbounded fluctuations do not arise from a Gaussian noise but from
a deterministic evolution like the ones in \cite{SE00,ESR03}: this is a simple 
extension of the main idea and method of \cite{CV03} and provides an
alternative interpretation to the analysis in \cite{SE00,ESR03}.

Our analysis can be applied to the example of the Anosov flow with singular 
metric considered above and to more realistic systems: among them
systems of particles interacting via an unbounded potential 
(like a Lennard--Jones (LJ) or a Weeks--Chandler--Andersen (WCA) 
potential), driven by an external field and subject to 
an isokinetic or a Nos\'e--Hoover thermostat. To be definite one
can consider system of $N$ particles in $d$ dimensions, described 
by evolution equations $\dot{\V p}_i=\V E-\dpr_{\V q_i}\Phi-\a \V p_i$,
$\dot{\V q_i}={\V p}_i$. For an isokinetic Gaussian thermostat, $\a$ is
a function of $\V p_i$, chosen so to keep the total kinetic energy fixed
to $\sum_i {\bf p}_i^2=Nd\b^{-1}$.
For a Nos\'e--Hoover thermostat $\a(t)$ is a variable independent of
$\V q_i(t),\V p_i(t)$ and satisfying the evolution equation $\dot\a=
\fra1Q\big[\sum_i\V p_i^2-Nd\b^{-1}\big]$, with $Q,\b>0$ parameters.

In both cases the phase space contraction $\s(x)$ has the form 
$\s_0(x)-\b\fra{d}{dt}V(x)$, 
where $\b$ has the interpretation of inverse temperature.
In the isokinetic case,
$\s_0(x)$ is bounded, and $V = \Phi$.
In the Nos\'e--Hoover case $\s_0(x)$ has, in the SRB distribution, 
a fast decaying tail
(Gaussian at equilibrium, and likely to
remain such in presence of external forcing)
and $V = \sum_i\fra{\V p_i^2}{2}+\Phi(\V q)+
Q\fra{\a^2}{2}$, \cite{N84,H85}.

In both cases, {\it in equilibrium}, the SRB probability of
$V$ has an exponential tail $\sim e^{-\b V}$ (possibly with power-law
corrections).
{\it For the purpose of illustration} we assume, from now on, that the
same happens in presence of the force $\V E$.
This is an {\it essential and far from obvious assumption} useful,
as discussed below, to
understand the possible role of the singularities, but it
should not be assumed lightly as it is well known that the SRB
distributions may have very peculiar $E$ dependence and, at the
moment, a not intuitive character, \cite{DLS02,BDGJL01}. Nevertheless, 
in preliminary numerical simulations, it seems approximately
correct, at least within the accuracy of the numerical data and for $|\V E|$
not too large; furthermore the analysis that follows can be naturally 
adapted to more general assumptions on the tails.

In such cases the non normalized
variable $a$ (introduced before Eq. (\ref{6})) 
has the form $a_0+\fra{\b}\t (V_i-V_f)$ where
$V_i,V_f$ are the values of $V(x)$ at the initial and
final instants of the time interval of size $\t$ on which $a$ is
defined, and $a_0 \Defi \fra1\t\ig_0^\t \s_0(S_tx)dt$:
\begin{equation}a=\fra1\t\ig_0^\t \s(S_tx)dt\=a_0+\fra\b\t (V_i-V_f)
\label{8}\end{equation}
If the system is chaotic and $\t$ is large, the variables $a_0, V_i,V_f$ can be
regarded as independently distributed, because $a_0$ depends 
essentially only on the length $\t$ of the time interval, while $V_i$ 
and $V_f$ depend on the precise locations of the extremes of the interval.
Moreover the distribution of $V=V_i$
or $V=V_f$ is essentially $\sim e^{-\b V} dV$ to leading
order as $V\to\io$, as discussed above. Therefore the 
rate function of the variable $a$ can be computed as
\begin{equation}\label{9}\begin{split}
&\lim_{\t\to\io}\fra1{\t}\log
\int_{- p^* \s_+}^{p^*\s_+} da_0 \int_0^\io dV_i 
\int_0^\io dV_f \, \cdot \\
&\cdot \, e^{\t \wt \z_0(a_0) - \b V_i -\b V_f} 
\d[\t(a-a_0)+\b V_i - \b V_f] \\
&=\lim_{\t\to\io}\fra1{\t}\log
\int_{- p^* \s_+}^{p^*\s_+} da_0 \, e^{\t\wt \z_0(a_0) - \t|a-a_0|}
\end{split}\end{equation}
where $\wt \z_0(a_0)$ is the rate function of $a_0$; thus
\begin{equation}\widetilde \z(a)=\max_{a_0 \in [-p^* \s_+,p^* \s_+]}
\Big[ \widetilde\z_0(a_0) - |a-a_0| \Big]\label{10}\end{equation}
Defining $a_\mp$ by $\widetilde \z_0 '(a_\mp) = \pm 1$, 
by the strict convexity of $\widetilde \z_0(a_0)$ it follows
\begin{equation}
\widetilde \z(a) = \left\{ 
\begin{array}{ll}
\widetilde \z_0(a_-) - a_- + a \ \ , & a < a_- \\
\widetilde \z_0(a) \ \ , & a \in [a_-,a_+] \\
\widetilde \z_0(a_+) + a_+ - a \ \ , & a > a_+ \\
\end{array} \right.\label{11}
\end{equation}
If we assume that $\wt\z_0(a_0)$ satisfies FR (as expected from the
chaotic hypothesis, see below), then $\widetilde \z_0(a_0) =
\widetilde \z_0(-a_0) + a_0$ and by differentiation it follows that
$a_-=-\s_+$, where $\s_+$ is the location of the maximum of
$\widetilde\z_0$, \ie is the average of $a$, and that
$\wt\z_0(a_-) = \wt\z_0(-\s_+)=\wt\z_0(\s_+)-\s_+=-\s_+$.
Moreover it is clear that
$a_+ > \s_+$ because $\wt\z'_0(a_+)<0$. Using these informations
one can show that, for $a \geq 0$:
\begin{equation}
\widetilde \z(a)-\wt\z(-a) = \left\{ 
\begin{array}{ll}
a \ \ , & a < \s_+ \\
\widetilde \z_0(a)+a \ \ , & \s_+ \leq a \leq a_+  \\
\widetilde \z_0(a_+) + a_+  \ \ , & a > a_+ \\
\end{array} \right.\label{12}
\end{equation}
It follows that, if $\wt\z_0(a_0)$ satisfies FR up to $a=p^*\s_+$, then  
$\widetilde \z(a)$ satisfies FR only in the
interval $|a|<|a_-|=\s_+$. Outside this interval $\widetilde\z(a)$
does not satisfy the FR and in particular for $a\ge a_+$ it is
$\widetilde\z(a)-\widetilde\z(-a)=const.$, as
already described in \cite{CV03}. Eq.~(\ref{12}) is the generalization
of the result of \cite{CV03} to the case where $\wt\z_0(a_0)$
is not Gaussian.

Translated into the normalized
variables $p_0=a_0/\s_+$ and $p = a/\s_+$, this means that,
even if the rate function of $p_0$ satisfies FR up to $p^*>1$, 
the rate function of $p$ verifies FR only for
$|p|\leq1$. This is the effect due to the presence of the singular
boundary term. Note that the scenario above applies only
to the case in which $V_i,V_f$ are unbounded and have exponential 
tails. A repetition of the discussion above in the case
that $V_i,V_f$ are unbounded but with tails faster than exponential
would lead to the conclusion that $\wt\z(a)=\wt\z_0(a)$. In particular if
$V_i,V_f$ are assumed to be bounded $\wt\z(a)=\wt\z_0(a)$. 
Of course in these cases the times of convergence of $\wt\z(a)$ to $\wt\z_0(a)$
will depend on the details of the tails of $V_i,V_f$ (for instance if $V_i,V_f$
are bounded by a constant $B$, the times of convergence will grow with $B$). 
Note also that the result above does not depend
on the details of the distribution of $V_i,V_f$ for small $V$ (in particular
it does not depend on the lower cutoff $V=0$ assumed in Eq.~\ref{9}).

\begin{figure}[t]
\includegraphics[width=.5\textwidth]{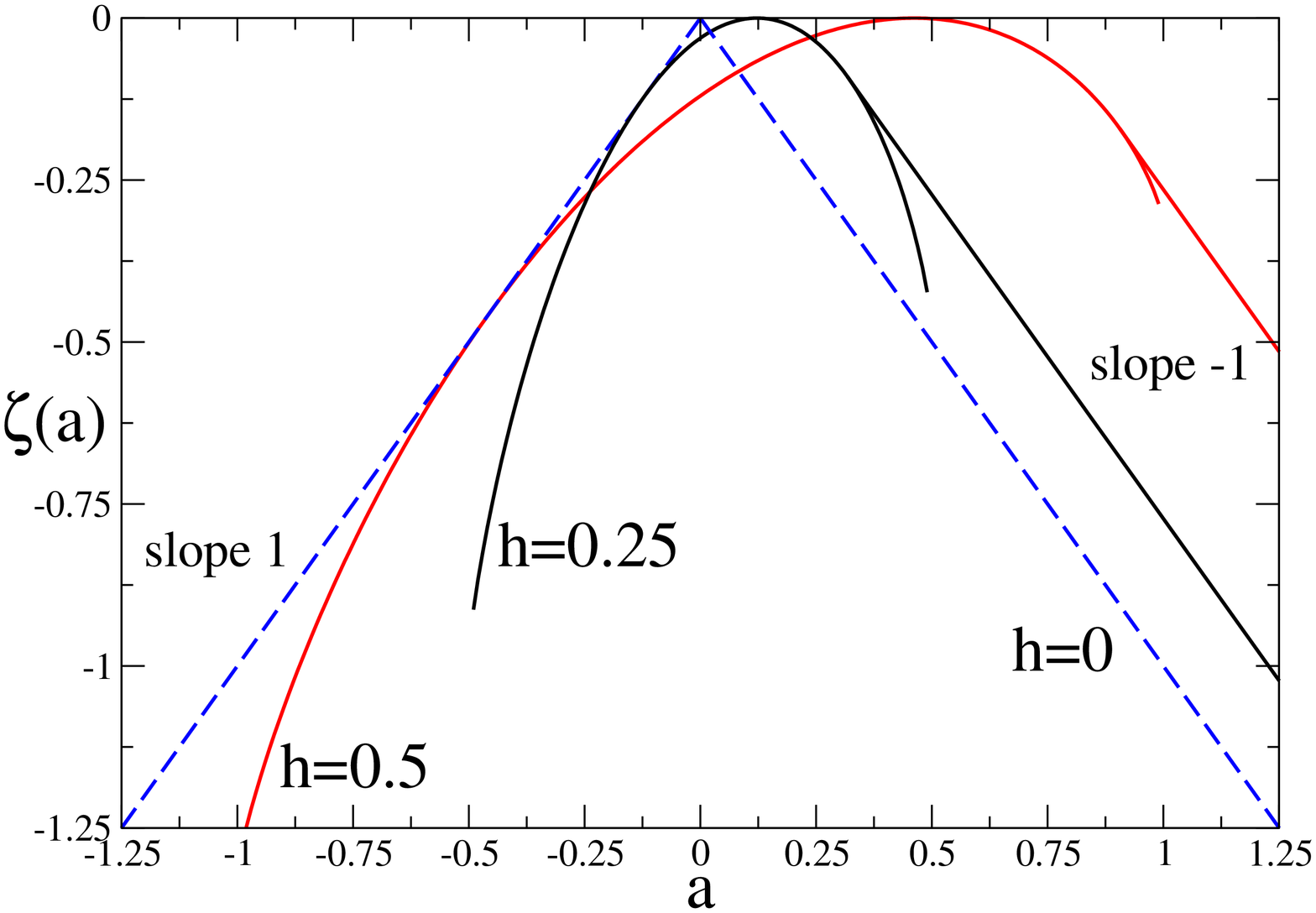}
\caption{An example in a stochastic model of FR.  The graph gives
the two functions $\wt\z_0(a)$ and $\wt\z(a)$ for $h=0.,0.25,0.5$.
The average of $a$ is $\media{a}=\s_+ = 2h \tanh h$,
$a_+ = 2h \tanh 3h$ and $a^* = 2 h$.  The
function $\wt\z(a)$ is obtained from $\wt\z_0(a)$ by continuing it
for $a<a_-=-\s_+$ and $a>a_+$ with straight lines of slope $\pm 1$. 
It does not satisfy the FR for $|a|>\media{a}$. As $h\to0$, $\media{a} \to 0$,
which means that the interval in which the FR is verified shrinks to $0$.
In this limit $a_+\to 0$, so $\wt \z(a)$ approaches $-|a|$ (dashed lines).
Rephrasing this in terms of
$p=\fra{a}{\media{a}}$ one obtains that FR remains always valid
for $|p|<1$, even as $h\to0$. The three curves for 
$\wt\z_0(a)$ have the same tangent
on left side. The function $\wt\z_0(a)$ is finite {\it only} in the
interval $[-2h,2h]$ and it is $-\infty$ outside it, while the function
$\wt\z(a)$ is finite for all $a$'a and is a straight line outside $[a_-,a_+]$.}
\label{fig_zeta}
\end{figure}

An example of $\widetilde \z(a)$ is reported in Fig.~\ref{fig_zeta}:
it is a simple stochastic model for the FT (taken from Sect. 5 in
\cite{BGG97}, see also the extensions in \cite{LS99,Ma99}).  The
example is the Ising model without interaction in a field $h$, \ie a
Bernoulli scheme with symbols $\pm$ with probabilities
$p_\pm=\frac{e^{\pm h}}{2\cosh h}$.  Defining
$a_0=\fra1\t\sum_{i=0}^{\t-1}2 h\s_i$, so that 
$\s_+ = \langle a_0 \rangle = 2h \tanh h$,
and setting $x\defi\fra{1+a_0/(2 h)}2$,
and $s(x)=-x\log x-(1-x) \log (1-x)$, one computes
$\wt\z_0(a_0)=s(x)+\fra12 a_0+ const$ which is {\it not Gaussian} and it is
defined in the interval $[-a^*,a^*]$ with $a^*=2 h$. In
this case the large deviation function $\wt\z_0(a_0)$ satisfies FR for
$|a_0|\le a^*$. If a singular term
$V=-\log (\sum_{i=0}^\io 2^{-i-1}\frac{\s_i+1}2)$ is added to $a_0$,
defining $a=a_0+\b(V_i-V_f)$
(with $\b=\log_2 (1+e^{2h})$ so that the probability distribution of $V$ is 
$\sim e^{-\b V}$ for large $V$),
the resulting $\wt \z(a)$ does not verify FR for
$a > \media{a} = 2h \tanh h$. In particular, for $h\to 0$, the interval
in which the FR is satisfied vanishes.

\subsection*{How to remove singularities}

From the discussion above it turns out that singular terms which are
proportional to total derivatives of unbounded functions (like the
term $\frac{dV}{dt}$ that appears in the phase space contraction rate
of thermostatted systems) can induce ``undesired'' (or ``unphysical'')
modifications of the large deviations function $\z(p)$.

On heuristic grounds, when dealing with singular systems, one could
follow the prescription that unbounded terms in $\s(x)$ which are
proportional to total derivatives should be {\it subtracted} from the
phase space contraction rate.  If the resulting $\s_0(x)$ is bounded (as
it is \eg for the Gaussian isokinetic thermostat models considered) 
or at least if the tails of its distribution decay faster than exponentially,
then its large deviations function should 
verify the FR for $|p|\leq p^*$, $p^*$ being the
intrinsic dynamic quantity defined above.

Note that after the subtraction of the divergent terms the remaining
contraction, in the considered cases, is bounded for isokinetic
thermostats or has a  tail decaying faster than
exponential in the case of Nos\'e--Hoover thermostats. 
In the following for definiteness
we will assume $\s_0$ bounded but the same discussion 
is valid for $\s_0$ unbounded with tails decaying faster than exponential.

If the singular terms are not subtracted, the FR {\it will appear to
be valid only for $|p|\leq 1$ even if $p^* > 1$}.  This seems to have
generated statements that the Chaotic Hypothesis does not apply to
isokinetic systems, see \cite{ESR03}.

The heuristic prescription above can be motivated by a careful analysis of
the proof of the fluctuation theorem for Anosov flows.
In the following let us call again $a$ the integral of the total phase space
contraction rate $\s(x)$ (which includes singular terms) and $a_0$ the integral
of the bounded variable $\s_0(x)$ from which singular total derivatives
have been removed.

The fluctuation theorem was proved in \cite{GC95,Ga96,Ru99} {\it for
Anosov maps} and only later it has been extended in \cite{Ge98} to Anosov
flows. Very sketchily, 
the extension of the 
fluctuation theorem to Anosov flows in \cite{Ge98} is proved 
as follows. One reduces the Anosov flow on $\O$ to a map via
a Poincar\'e's section, associated with surfaces on $\O$ transversal
to the flow. The passage of the flow through any one of such
surfaces is called a {\it timing event}. The map between two consecutive
timing events is called a ``Poincar\'e's map''. The union $\O_P$ of the
surfaces represents the phase space of the Poincar\'e's map.
The surfaces in $\O_P$ can be suitably chosen, in such a way that the 
Poincar\'e's map is a chaotic map which although not smooth, hence not
an Anosov map, has (a non trivial fact \cite{Ge98}), 
all the properties necessary to prove the
fluctuation theorem (which therefore applies to systems
more general than the Anosov maps, although there is not a
general characterization of the systems which are not Anosov and to
which it applies). So, for such a map the fluctuation 
theorem holds and this in turn leads to a FR for the flow by the 
theory in \cite{Ge98} {\it under the assumption that the variable $\s(x)$ 
is bounded}. 

If, as in the case under analysis, $\s(x)$ is not bounded,
we can interpret the chaotic hypothesis as applying to the map
associated with a Poincar\'e's section which avoids the singularities
of the potential, a very natural prescription which allows us to
apply the theory in \cite{Ge98} and derive a FR for both the map and
the flow. For instance, we can choose as
timing events the instants in which either the potential energy or the 
Nos\'e's ``extended Hamiltonian'' exceed some fixed value $\bar V$. 
If we make this choice, the (discrete) average $\hat a$ of $\s(x)$ over a 
sequence of iterations of the Poincar\'e's map will coincide with 
the (discrete) average $\hat a_0$ of $\s_0(x)$ along the same sequence: 
this simply follows from the remark that by construction the total increment 
of $\s(x)-\s_0(x)$ between two timing events, given by $\b (V_f-V_i)$, is $0$ 
(by construction $\O_P$ 
is chosen as a subset of $\{x\in\O\,:\, V(x)=\bar V\}$ 
where $V_f=V_i$). Then, by the same 
argument in \cite{Ge98}, the fact that the rate function of $\hat a_0$ 
satisfies a FR and that $\s_0(x)$ is bounded implies that the rate function 
of the continuous average $a_0$ of $\s_0(x)$ along a trajectory of the flow
will satisfy the fluctuation theorem.

Therefore the distribution of $a_0$ will satisfy the FR
({\it by the chaotic hypothesis}) for $|a_0|<p^* \s_+$.  By the above
maximum argument, the distribution of $a$ {\it will also verify, as
a consequence}, the FR but only for $|a|\leq \s_+$, \ie in the form
(\ref{12}).

Then the (natural) prescription to study FR for chaotic flows is
to reduce the problem to a chaotic map considering only Poincar\'e's
sections which do not pass through a singularity of $\s(x)$. The
sum of $\s(x)$ over a large number of timing events on such
sections is equal to the time integral of $\s_0(x)$ plus a bounded
term which can be neglected. Thus the prescription on the choice of 
Poincar\'e's sections is equivalent to the heuristic prescription of 
removing from $\s(x)$ all the unbounded total derivatives.

It follows that the chaotic hypothesis leads to a clear prediction on
the outcome of possible numerical simulations of particle systems interacting 
via unbounded potentials and subject to the isokinetic
or the Nos\'e--Hoover thermostat: the FR will hold for all $|a|\leq \s_+$ and,
once the term $\frac{dV}{dt}$ is removed, for all $|a_0| < p^* \s_+$
with $p^* \geq 1$. Note that in the cases under analysis $a_0$ coincides with
the dissipation function of Evans and Searles that was in fact predicted to 
satisfy FR \cite{SE00,ESR03}, even though for different reasons. 
We believe that 
the correct interpretation of the fact that FR for $\wt\z_0(a_0)$ holds 
for all $|a_0| < p^* \s_+$ is the one given above.

The numerical results of \cite{DK04,ZRA03,GZG05} agree with 
the prediction that 
FR for the rate function of $a_0$ is valid even beyond $a_0=\s_+$. 
The prediction that (at least near equilibrium) the rate function of $a$
should satisfy FR only up to $a=\s_+$ and that should become linear for $a\ge
a_+$ at the moment has been experimentally confirmed only in Gaussian cases 
\cite{CV03a,CV03,VCC04}. 
It would be very interesting to investigate in detail 
the structure of $\wt\z(a)$ even in non Gaussian cases. Note that this is 
far from being an easy task (in particular the analysis in \cite{GZG05}
was not sophisticated enough to study this problem). 
In fact, as discussed in detail in \cite{ZRA03}, 
the presence in the definition of $\s$ of a total derivative of an unbounded 
function may enlarge of $2$ orders of magnitudes the times needed for the 
probability distribution of $a$ to reach its asymptotic shape: even 
in the Gaussian region (small fluctuations of $a$ around $\s_+$)
the convergence times for $\wt\z(a)$ are found to be
of order $1000$ decorrelation times, 
versus a time of order $10$ decorrelation times needed for $\wt\z_0(a_0)$ 
to converge to its asymptotic shape~\cite{ZRA03}. Clearly, for times of order 
$1000$ decorrelation times, it is very hard to observe 
fluctuations of $a$ larger than $a_+-\s_+$. In order to verify the prediction 
for the shape of $\wt\z(a)$ beyond $a=a_+$ an experiment 
specifically designed for this purpose would be 
needed, together with a detailed investigation of the finite time corrections 
to $\wt\z(a)$, along the lines in \cite{GZG05}.

\section{Conclusions and remarks}
\label{sec:III}

We showed that the Chaotic Hypothesis can be applied even to singular 
chaotic systems (in particular even
to Gaussian isokinetic or Nos\'e--Hoover thermostatted systems), 
by identifying their macroscopic behavior with
that of reversible Anosov systems with singular metric. 
Reversible Anosov systems with singular metric are systems to which the 
mathematical analysis usually leading to FR can be (rigorously) repeated 
to lead to a modified FR, illustrated by Eq.s~(\ref{11}), (\ref{12}). Note that 
for $\s_+\to 0$ Eq.~(\ref{11}) tends to the distribution $\wt\z(a)=-|a|$ 
violating the usual FR (simply because the limiting distribution is 
symmetric, by time reversal).
For Anosov systems with singular metric, the prescription to 
avoid oddities (\ie to avoid a modified FR) is to subtract 
from $\s(x)$ a total derivative $dV/dt$, in such a way that the variable
$\s_0=\s-dV/dt$ is bounded or has faster than exponential tails. 
The distribution of $\s_0$ will verify the
FR also for $|p|>1$. This prescription is equivalent
to the very reasonable prescription that the Poincar\'e's section used for
mapping the flow into a map does not pass through a singularity of $\s(x)$.
Accepting the Chaotic Hypothesis, we propose to apply the 
same prescription to remove singularities to singular chaotic systems. 
Our prescription coincides with other prescriptions proposed earlier 
(for different reasons) in the literature. \\
The analysis in Sect. 2 above applies as well to understand how to
apply the FR to systems with Gaussian (or unbounded) noise and the
compatibility between the general theory of \cite{Ku98,LS99,Ma99}
with the works \cite{CV03} and \cite{VCC04,GC05}. \\
\\
Note that the picture we propose is different from the interpretation
of the apparent violations to FR in singular systems proposed recently
in \cite{ESR03}, where in particular
it is argued that FR and CH do not hold for thermostatted systems near 
equilibrium. We conclude by comparing more closely our discussion 
with the corresponding discussion in \cite{ESR03}.\\
\\
\0(1) As stressed above it is dangerous (and wrong) to consider (\ref{6}) 
without the restriction $|a|\le p^*\s_+$ as the prediction of fluctuation 
theorem. In \cite{ESR03} the authors, after having correctly
pointed out this point, seem (quite surprisingly!) to forget about
this condition in the following.  
For instance, when studying the problem of approach to equilibrium, 
in order to show a contradiction between FR and GK relations, they
assume that the relation Eq.(\ref{6}) without the condition $|a|<p^*\s_+$
``is correct both at equilibrium and near equilibrium'' and they proceed 
to infer from this a contradiction. Of course such assumption is wrong 
and the fact that from this contradictions follow is not an argument 
against CH or FR.

\0(2) An argument in \cite{ESR03}
is supposed to prove that the relation $\widetilde\z(-a)=\widetilde\z(a)-a$
(without the condition $|a|<p^*\s_+$) holds for reversible Anosov
systems for all $a$'s, also for $\s_+=0$ (in particular they say that 
``{\it the division by $\s_+$ does not seem to be necessary for the
proof in [37]}''). This is not the case: 
at equilibrium as well as near equilibrium, as remarked above and as
illustrated also by \cite{BGM98}, there are examples of systems
for which the proof of FR can be {\it rigorously} repeated step by step 
but for which the correct conclusion of the proof
is that the relation $\widetilde\z(-a)=\widetilde\z(a)-a$
is violated for $a> p^*\s_+$. For instance, this is the case for a 
conservative Anosov flow with singular metric (in which the relation above 
is violated trivially by time reversal).
These counterexamples show that the assumption 
$\s_+>0$ {\it is, instead, essential} for the proof of fluctuation theorem. 
The necessity of the assumption $\s_+>0$ is stressed in
the early paper \cite{Ga95b} which the Authors of \cite{ESR03} quote; 
it is stressed also in the paper \cite{Ru99} which also makes clear that
$\widetilde\z(-a)=\widetilde\z(a)-a$ can only hold under the
assumption that $|a|$ does not exceed a maximum value.

\0(3) The analysis in Sect. 2 above shows that the probability
distribution describing isokinetic systems near equilibrium {\it
are SRB distributions} (contrary to what is claimed in \cite{ESR03}):
this is mathematically obvious by the very definition of SRB
distribution in the case of Anosov systems ({\it even if isokinetic}
or in general with singular metric, see \eg 
the geodesic flow discussed above)
and it appears to be true also in non Anosov systems that
have so far been considered.

\0(4) In the case of the thermostatted particle systems considered in 
\cite{ESR03} the unbounded derivative $\frac{dV}{dt}$ is also the
contraction rate of the volume in equilibrium, \ie with ${\bf E}=0$. 
Thus, for $\V E\neq \V 0$,
one can remove the total derivative from $\s(x)$ simply considering
the contraction {\it with respect to the equilibrium invariant
distribution} $e^{-\b V}$, as stated in \cite{ESR03, SE00}. 
However, this
observation does not provide a general prescription to remove the
singular part from the phase space contraction rate because it rests on
the very special fact that the singularities of the function $V(x)$
(\ie of the potential $\Phi$) do not depend on ${\bf E}$. 
The prescription that the phase space contraction should be computed
on non singular Poincar\'e's sections, instead, does not require
any other assumption. In general the two prescriptions and 
the corresponding predictions differ and we believe that in general
the prescription of computing $\s$ with respect to the equilibrium invariant
distribution has not the desired effect of removing all singularities
(then in general $\s$ with respect to the equilibrium invariant
distribution could violate FR for $\s_+<a<p^*\s_+$).

\acknowledgments

We are indebted to E.G.D. Cohen, and R. Van Zon for many enlightening
discussions. F.Z. wish to thank G.Ruocco for many useful discussions.
G.G is indebted to Rutgers University, I.H.E.S and 
\'Ecole Normale Sup\'erieure, where he spent periods of leave while
working on this project.

\bibliography{nth2} 

\bibliographystyle{apsrev} 
\*

\def\revtexz{{\bf
R\lower1mm\hbox{E}V\lower1mm\hbox{T}E\lower1mm\hbox{X}}} 

\0e-mail: {\tt 
bonetto@math.gatech.edu\\
giovanni.gallavotti@roma1.infn.it\\
alessandro.giuliani@roma1.infn.it\\
francesco.zamponi@phys.uniroma1.it}\\
web: {\tt http://ipparco.roma1.infn.it}\\
Dip. Fisica, U. Roma 1,\\
00185, Roma, Italia

\revtex
\end{document}